\documentclass[twocolumn,showpacs,preprintnumbers,amsmath,amssymb]{revtex4}

\usepackage{graphicx}
\usepackage{dcolumn}
\usepackage{bm}

\begin{document}

\preprint{in preparation for Physical Review Letters}

\title{
Collective single mode precession of electron spins in a quantum
dot ensemble}

\author{A. Greilich$^1$, S. Spatzek$^1$, I. A. Yugova$^2$, I. A. Akimov$^{1,3}$, D.~R. Yakovlev$^{1,3}$,
Al.~L. Efros$^4$, D. Reuter$^5$, A.~D. Wieck$^5$, and M. Bayer$^1$}
\affiliation{$^1$ Experimentelle Physik 2,
                Technische Universit\"at Dortmund,
                D-44221 Dortmund, Germany}
\affiliation{$^2$ Institute of Physics,
                St. Petersburg State University,
                198504 St. Petersburg, Russia}
\affiliation{$^3$ A. F. Ioffe Physico-Technical Institute, Russian
Academy of Sciences, 194021 St. Petersburg, Russia}
\affiliation{$^4$ Naval Research Laboratory,
                20375 Washington DC, USA}
\affiliation{$^5$ Angewandte Festk\"orperphysik,
             Ruhr-Universit\"at Bochum,
             D-44780 Bochum,
             Germany}

\date{\today}

\begin{abstract}
We show that the spins of {\sl all} electrons, each confined in a
quantum dot of an (In,Ga)As/GaAs dot ensemble, can be driven into
a single mode of precession about a magnetic field. This regime is
achieved by allowing only a single mode within the electron spin
precession spectrum of the ensemble to be synchronized with a
train of periodic optical excitation pulses. Under this condition
a nuclei induced frequency focusing leads to a shift of all spin
precession frequencies into the synchronized mode. The macroscopic
magnetic moment of the electron spins that is created in this
regime precesses without dephasing.
\end{abstract}

\pacs{78.67.Hc, 78.55.Cr}

\maketitle

Solid state implementations of quantum information processing
promise scalability towards large numbers of
qubits~\cite{loss_pra98,QBbook}. However, they are typically
impeded by the non-ideal crystal environment leading to a wide
dispersion of the properties of elementary excitations envisaged
as qubits. This gives rise to a number of complications: a single
excitation needs to be isolated, which often requires high
resolution in space, energy, etc. The read-out signal of such an
excitation is typically weak, so that measurements may require
times comparable to the decoherence
time~\cite{atatuere_natphys07}. These problems may be overcome, if
one had access to an ensemble of identical quantum bits, all
prepared in the same quantum state, which is, however, complicated
by the unavoidable inhomogeneities.

A qubit candidate with promising features is an electron spin
confined in a quantum dot (QD)
~\cite{loss_pra98,QBbook,atatuere_natphys07,gammon-still,elzerman_04,kroutvar_04,imamoglu_prl99,wolf_science01}.
Its decoherence time $T_2$ at cryogenic temperatures, for example,
is in the microseconds range, as determined by a spin-echo
measurement on a single GaAs/AlGaAs gated
QD~\cite{petta_science05}. This property, which should allow one
to perform many operations coherently, is, however, obscured in a
QD ensemble by fast dephasing of electron spin polarization due to
frequency dispersion for precession about a transverse magnetic
field~\cite{merkulov_prb02,awschalom_02}. The dephasing could be
suppressed for particular electron spin subsets by synchronizing
their precession with the repetition rate of the periodically
pulsed laser used for generation of spin
polarization~\cite{greilich_science06}. The precession frequencies
in these subsets satisfy the mode-locking condition:
$\omega_K=2\pi K/T_R$, where $T_R$ is the pulse repetition period
and $K$ is an integer. Fulfillment of this condition gives rise to
bursts in the Faraday rotation (FR) signal measured from a
(In,Ga)As/GaAs QD ensemble right before excitation pulse arrival.
The FR signal decay allowed us to measure $T_2=3$~$\mu$s
 ~\cite{greilich_science06}.

The majority of electrons in the ensemble would not satisfy the
mode-locking condition, if the electron spin precession frequency
in an individual dot was determined just by the external magnetic
field $B$ and the electron g-factor $g_e$. In most III-V compound
QDs, however, an electron is also exposed to the collective
hyperfine field of the dot nuclei. As a result, the electron spin
precession frequency, $\omega=\mu_Bg_eB/\hbar+\omega_{N,x}$,
contains the nuclear contribution, $\omega_{N,x}$, which is
proportional to the projection of the nuclear spin polarization on
the external field ($\mbox{\boldmath $B$}\|\mbox{\boldmath $x$}$).
Here $\mu_B$ is the Bohr magneton. The magnetic field suppresses
magneto-dipole interactions between nuclei and, in darkness, the
projection of the nuclear spin polarization does not change for
hours or even days. The resonant optical excitation of the QDs
leads, however, to light-assisted flip-flop processes between
electron and nuclei. The consequent random fluctuation of
$\omega_{N,x}$ eventually drives almost all electron spins in the
ensemble into mode-locked modes, corresponding to a nuclear
induced frequency focusing effect~\cite{greilich_science07}. The
laser protocol applied in Ref.~\cite{greilich_science07} still
leads to precession on a considerable number of different modes.
Therefore, the FR traces show dephasing of spin coherence on a
ns-time scale. The complete suppression of the dephasing would
require precession of all electron spins in the QD ensemble on a
single frequency, i.e. focusing of the spins to a single mode.

In this paper we demonstrate the achievement of an almost pure
single precession mode regime in an ensemble of $\sim$ a million
QDs via proper tailoring of the laser excitation protocol and the
magnetic field strength. This regime allows controlled switching
between single and double mode precession in the magnetic field
range from about 50\,mT to 1\,T.

The time-resolved pump-probe Faraday rotation measurements were
performed on singly negatively charged (In,Ga)As/GaAs
self-assembled QDs (see Ref.~\cite{greilich_prl} for details). 
The sample was held at a temperature $T=6$~K in a superconducting
split coil and magnetic fields were applied perpendicular to the
sample growth axis. For optical excitation we used a mode-locked
Ti:Sapphire laser emitting 4~ps pulses at a rate of 75.6\,MHz
(13.2\,ns pulse separation). After splitting the laser beam with a
50:50 intensity ratio, one beam was sent over a delay line to be
retarded by 6.6\,ns. After reuniting the split pulse trains along
the optical axis, the pump pulse repetition rate was doubled to
151.2~MHz, corresponding to $T_R = 6.6$\,ns. The photon energy was
tuned to the QD ground state optical transition. Circular
polarization  of the pump pulses was modulated at 50 kHz. Probe
pulses were linearly polarized and their energy was either equal
to the pump  (degenerate regime) or different from it
(nondegenerate pump-probe).

\begin{figure}[htb]
\centering
\includegraphics[width=6.5cm]{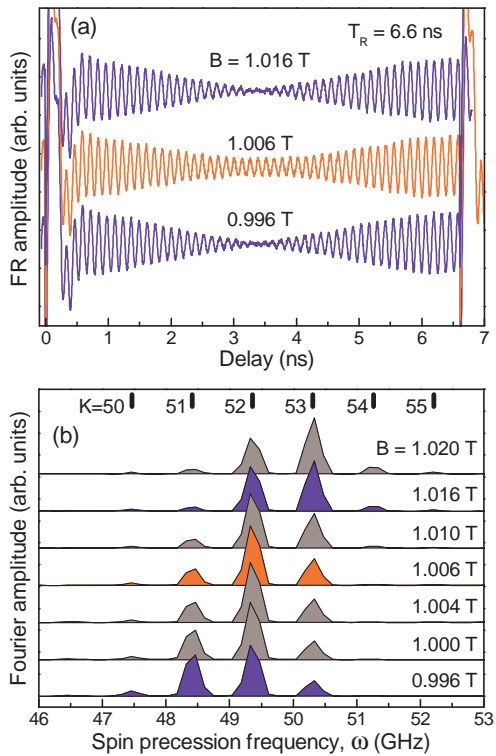}
\caption{(color online) (a) Faraday rotation spectra recorded
around $B$ = 1\,T. Pump and probe with degenerate photon energies
at 1.38~eV have powers of 23 and 13~W/cm$^2$, respectively. (b)
Fourier transforms extracted from FR signals measured over an
eight period time interval of 54~ns.  Positions of
phase-synchronized modes $\omega_K$ are marked by vertical
lines.}\label{fig:1}
\end{figure}

Figure~\ref{fig:1}(a) shows Faraday rotation traces for magnetic
fields around 1\,T. The pump pulses hit the sample at times $t=0$
and 6.6\,ns. Over the narrow  range of magnetic fields the FR
traces undergo strong modifications although the oscillation
frequencies appearing in them do not change significantly. At
$B=0.996$ and 1.016\,T the signal amplitude shows a strong decay
after the first pulse, hits a node in the middle between the pumps
and afterwards increases symmetrically towards the second pulse.
However, at the intermediate field of 1.006~T, deviating by 10~mT
only from the two other traces, the signal decay after the first
pulse is weaker and, in particular, it does not show a node. This
non-monotonic behavior of the FR signal with $B$ is repeated every
0.02~T and is typical for signals involving only a few precession
modes.

Indeed, the Fourier spectra in Fig.~~\ref{fig:1}(b) confirm that
the FR signal in this interval of magnetic fields is contributed
by 4--6 electron spin precession modes, out of which 2--3 have
strong weight. These Fourier transforms were obtained by
integrating over a time interval equal to eight repetition periods
$T_R$ to improve the resolution. With increasing magnetic field
the center of the contributing precession frequencies is seen to
shift to higher values relative to discrete mode-locked frequency
spectrum, $\omega_K=2\pi K/T_R$, where $K=50$--55 around 1 T. This
shift changes the relative contribution of the different modes to
the Fourier spectra. If for a magnetic field two strong modes of
the same weight dominate the spectrum, the FR signal shows a node
between the pump pulses (see $B=0.996$ and 1.016~T). For the
intermediate field of $B=1.006$~T, the FR signal is determined by
a central mode which is accompanied, however, by strong symmetric
satellites. Therefore, the FR signal has a considerable amplitude
in between the pulses.

\begin{figure}[htb]
\centering
\includegraphics[width=\columnwidth]{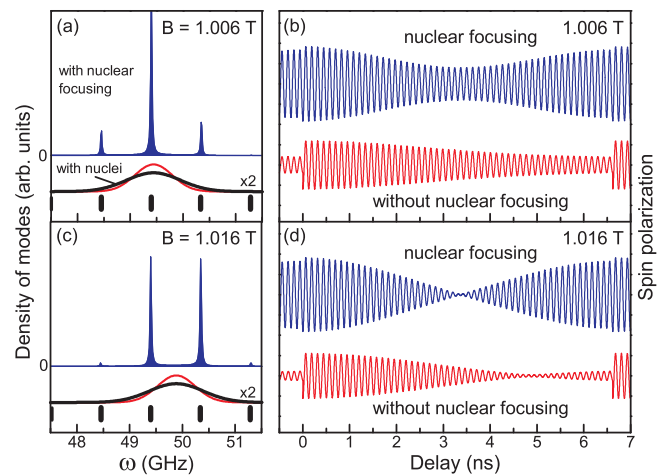}
\caption{(color online) Calculations of the density of electron
spin precession modes $\rho(\omega)$ and corresponding FR signals
at $B=1.006$ and $1.016$\,T. Panels (a),(c) show the density of
modes created by $g$-factor dispersion (red line), by nuclear
fluctuations (black line) and by nuclear induced frequency
focusing. Positions of phase-synchronized modes $\omega_K$ are
marked by vertical lines. Panels (b),(d) show the spin
polarization which is proportional to the FR signal magnitude,
calculated with and without frequency focusing. Parameters for
calculations: $g_e=0.556$,  $\Delta g_e=0.004$,
$\Delta\omega_{N,x}=0.37$~GHz, and pump pulses with  $\pi$-area
~\cite{petrov_prb}.}\label{fig:3}
\end{figure}

These experimental results are well described by the model of
nuclear induced frequency focusing~\cite{greilich_science07}.
Under resonant optical excitation the nuclei drive electron spin
precession frequencies in all dots to modes satisfying the phase
synchronization condition (PSC): $\omega_K=2\pi K/T_R$. This
redistribution removes the non-synchronized background in the
precession mode density, $\rho(\omega)$, by focusing to a few
leading modes. In Fig.~2 we show the theoretical mode density and
the corresponding Faraday rotation spectra created by the train of
$\pi$ pulses with repetition period $T_R=6.6$\,ns at $B=1.006$ and
1.016\,T.  Calculations were conducted using the procedure
described in Ref. \cite{greilich_science07} for the parameter set
described in figure caption. These results are in good agreement
with the experimental data. For comparison we also show the
corresponding dependencies without nuclear frequency focusing. The
significant deviation from the experimental observations
underlines the importance of the nuclear contribution in the
mode-locking.

\begin{figure}[htb]
\centering
\includegraphics[width=7cm]{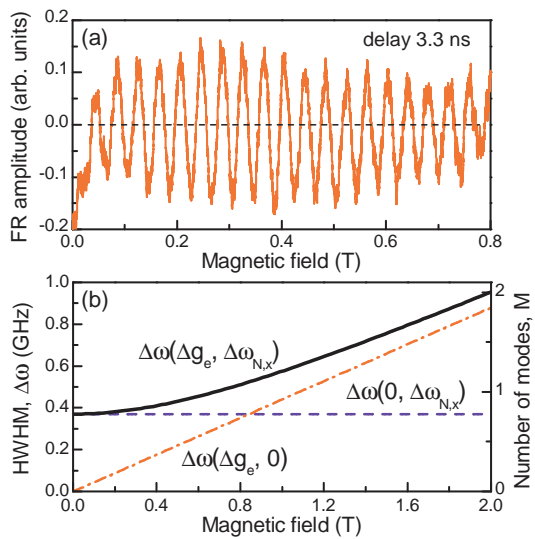}
\caption{(color online) (a) Magnetic field dependence of the FR
signal at 3.3~ns delay for pump and probe powers of 13~W/cm$^2$
each. (b) Magnetic field dependence of dispersion of electron spin
precession frequencies, $\Delta\omega(\Delta
g_e,\Delta\omega_{N,x})$, (left scale) and of number of
mode-locked frequencies (right scale) contributing to the FR
signal (solid line). Dashed line shows the nuclear contribution to
the dispersion, $\Delta\omega_{N,x}$, in our QDs. The contribution
by the electron spin $g$-factor dispersion is shown by dash-dotted
line. Parameters are the same as in Fig.~2.} \label{fig:2}
\end{figure}

The experimental data suggest that the magnitude of the FR signal
in the mid between the pump pulses (at a delay of 3.3\,ns) is a
characteristic quantity of the precession mode structure. The
magnetic field dependence of this magnitude in Fig.~\ref{fig:2}(a)
oscillates symmetrically around zero value. The oscillations are
connected with alternating dominance of either an odd or an even
number of modes contributing to the FR signal. The strongest
signal, either minimum or maximum, is reached for an odd number of
modes, while zero signal is caused for an even mode number. This
sequence makes an impression that a single mode regime could be
ultimately reached just by decreasing the magnetic field strength.

Let us  estimate, however, the number of mode-locked modes, which
contribute to the FR signal for a particular magnetic field
strength. The separation between these modes is $2 \pi / T_R$
\cite{greilich_science06}. Consequently, the number of such modes,
$M$, is:
\begin{equation}\label{eq:1}
M = 2\Delta \omega(\Delta g_e,\Delta\omega_{N,x})T_R/2\pi,
\end{equation}
where $\Delta \omega(\Delta g_e,\Delta\omega_{N,x})$ is the
half-width dispersion of electron spin precession frequencies in
the QD ensemble:
\begin{equation}\label{eq:2}
\Delta \omega (\Delta g_e,\Delta\omega_{N,x}) = \sqrt{[\mu_B\Delta g_e B/\hbar]^2 + \Delta
\omega_{N,x}^2}~.
\end{equation}
Here $\Delta g_e$  is the dispersion of electron $g$-factors in
the ensemble of optically excited dots and $\Delta \omega_{N,x}$
is the nuclear contribution to the dispersion of electron spin
precession frequencies in each dot. The magnitude of $\Delta
\omega_{N,x}$ is determined by statistical fluctuations of the
nuclear spin polarization projection onto the magnetic field in
the dot volume \cite{merkulov_prb02}. For our dots $\Delta
\omega_{N,x}=0.37$\,GHz~\cite{greilich_science07}.

Equations (\ref{eq:1},\ref{eq:2}) define a clear strategy for
achieving the single mode precession regime in a QD ensemble. The
number of mode-locked modes can be reduced: (a) by minimizing
$\Delta g_e$, (b) by decreasing $B$, and (c) by reducing $T_R$.
Each of these approaches has, however, its own restriction.

(a) Generally, the dispersion $\Delta g_e$ in a QD ensemble is
connected to variations of  dot shape and size. For (In,Ga)As dots
a systematic dependence of the electron $g$ factor on energy of
the band edge optical transitions has been observed
\cite{greilich_prl}. $\Delta g_e$ can then be controlled by the
laser spectral width, which is inversely proportional to the pulse
duration. However, as one is interested in fast spin
initialization, the duration should not exceed $\sim$ 10\,ps
(spectral width $\sim 0.1$~meV). This is an important limitation
because the efficiency of spin polarization initialization drops
when the pulse duration becomes comparable with the times of hole
spin precession and electron-hole
recombination~\cite{greilich_prl}.

(b) A reduction of the magnetic field strength is possible to an
extent that it is still considerably larger than the randomly
oriented effective field of the nuclei. Otherwise the nuclei would
induce fast dephasing~\cite{khaetskii_prl02,merkulov_prb02}. In
our dots the random nuclear fluctuation field has an amplitude of
about 7.5\,mT~\cite{yugova_prb07}. Also, to conserve nuclear spin
polarization, which is needed for frequency focusing, the magnetic
field should exceed the hyperfine field of the electron acting on
the nuclei (Knight field), which
is about 1 -- 3\,mT in our dots.

(c) A reduction of the repetition period is generally limited by
the trion decay time and the time scale, which multiple coherent
operations would require. For practical reasons, to observe the
single mode regime, $T_R$ should be longer than the ensemble
dephasing time of a few ns~\cite{greilich_prl}.

Figure ~\ref{fig:2}(b) shows the magnetic field dependence of
precession frequency dispersion $\Delta \omega(\Delta
g_e,\Delta\omega_{N,x})$ and of the mode number $M$ calculated for
our dots under the applied experimental conditions. For $B <$
0.8\,T, we estimate $0.75 < M \leq 1$, giving a lower limit for
the number of mode-locked frequencies in the FR signal. In this
range the $\Delta g_e$ contribution to $\Delta \omega$
(dash-dotted line) is smaller than the nuclear induced dispersion
(dashed line). The total dispersion is approximately equal to
$\Delta\omega_{N,x}$, which is independent of magnetic field. 
This results in periodic switching between almost pure single and
double mode regimes for $B <$ 0.8\,T, as
shown in Fig.~\ref{fig:2}(a). 
Whether one or two modes fall within the dispersion can be
adjusted by the magnetic field, which shifts the spectrum of
optically excited electron spin precession frequencies in the QD
ensemble relative to the spectrum of phase synchronized modes.
Magnetic fields larger than  0.8\,T increase the dispersion and
allow $\Delta \omega$ to cover more than three mode-locked
frequencies. This increases the amplitude of side modes
significantly, as seen in Fig.~1(b) for $B= 1$\,T,  and,
consequently, leads to considerable dephasing.

\begin{figure}[htb]
\centering
\includegraphics[width=7.5cm]{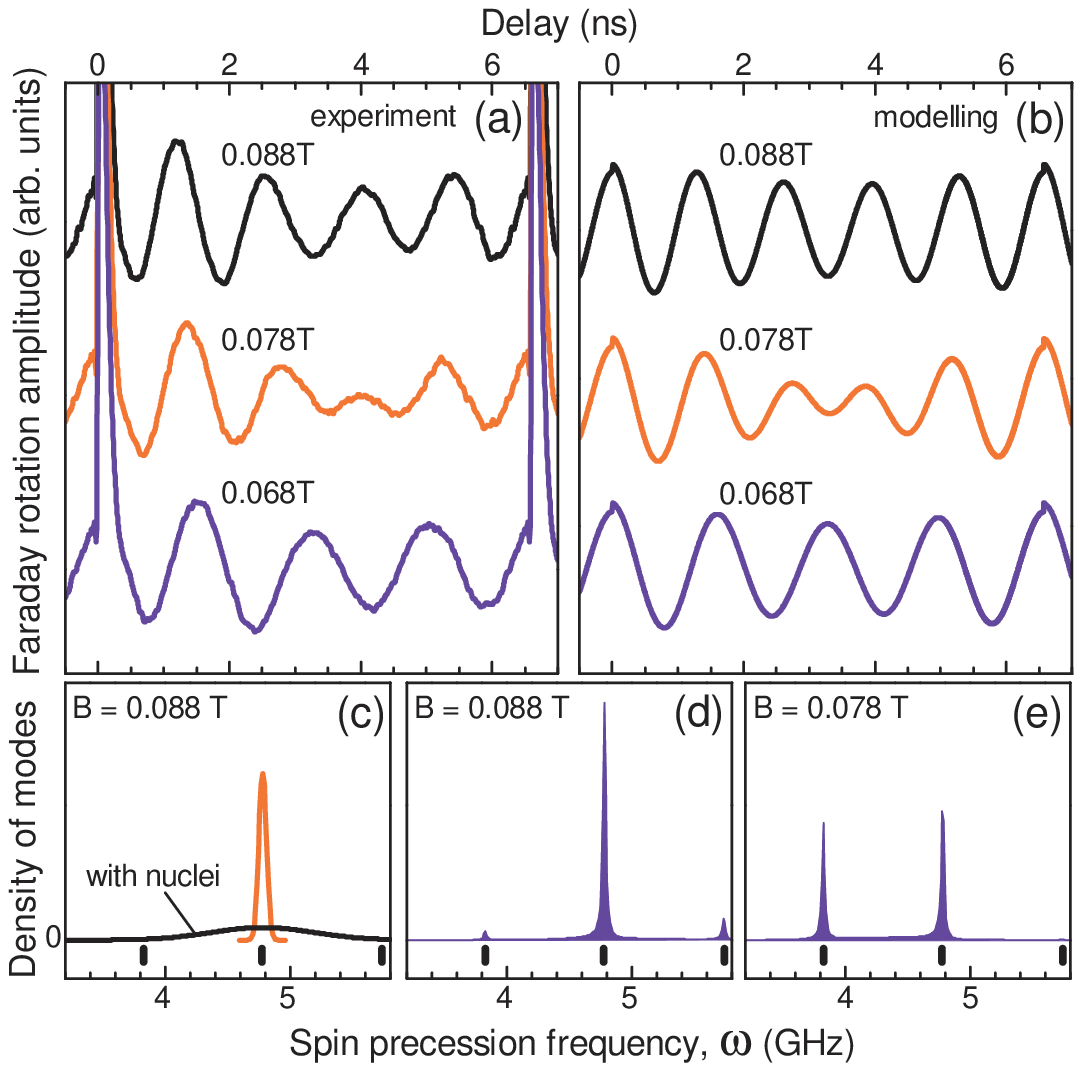}
\caption{(color online). (a) FR traces measured at weak magnetic
fields, to demonstrate the switching from single-mode to
double-mode regime. Pump energy 1.3837~eV at power 130~W/cm$^2$,
probe energy 1.3842~eV at power 13~W/cm$^2$. (b) Modelled FR
signals. (c) Calculated density of electron spin precession modes
due to g-factor dispersion (red line) and nuclear fluctuations
(black). (d,e) Calculated density of modes accounting for the
nuclear induced frequency focusing. Parameters are the same as in
Fig.~2. Positions of phase-synchronized modes $\omega_K$ are
marked by vertical lines.}\label{fig:4}
\end{figure}

We addressed experimentally the single mode regime by applying
very weak magnetic fields in order to minimize the contribution of
$\Delta g_e$. The experimental spectra shown in Fig.~4(a) were
measured around 0.078~T by nondegenerate pump-probe to increase
the signal contrast. A spectrum with a node in the middle is seen
at 0.078~T. However, for fields of 0.068 and 0.088~T, the FR
amplitude shows virtually no decay between the pump pulses.
Indeed, the calculated mode density in Fig.~4(d) is dominated by a
strong central peak at 4.78~GHz (corresponding to $K=5$). Two
satellites, which arise from the small overlap of $\rho(\omega)$
with the three phase synchronized modes [see panel (c)], are
hardly visible. It is remarkable that under this condition about
90\% of precession frequencies are focused to the single mode.


The single mode behavior of the QD ensemble in weak magnetic
fields is slightly obscured by quite weak dephasing due to the
very weak side modes in the precession frequency spectrum.
 For a true single
mode regime the condition: $\Delta\omega_{N,x}< 2\pi/T_R$ should
be satisfied. This can be reached by increasing the dot size
because $\Delta\omega_{N,x}$ is controlled by the statistical
fluctuations of the nuclear spin polarization in the dot volume,
$V$, as given by $\Delta\omega_{N,x}\sim 1/\sqrt{V}$. Otherwise,
quantum dots with other nuclear composition, e.g. with nuclei
having zero spin,  might be studied.

In summary, we have demonstrated that mode-locking combined with
nuclear induced frequency focusing allows us to drive an entire
ensemble of electron spins, confined in singly charged quantum
dots, into coherent single mode precession. This regime was
reached for relatively weak magnetic fields ($<0.6$\,T) and a
laser repetition period of 6.6\,ns.  The coherently synchronized
precession of a million spins in the QD ensemble represents a
macroscopic magnetic moment, which can be considered as
macroscopic quantum bit. Therefore, this regime will be very
useful for studying various coherent phenomena, such as
electromagnetically induced transparency or control-NOT gate
operations.

We acknowledge the support by the BMBF project 'nanoquit' and the
Deutsche Forschungsgemeinschaft (SPP1285). A.L.E. acknowledges
support of the Office of Naval Research and Alexander-von-Humboldt
Foundation.

\end{document}